\documentclass[12pt]{iopart}

\usepackage[english]{babel}
\usepackage{amsbsy}
\usepackage{amstext}
\usepackage{pst-node}
\usepackage{graphicx}
\usepackage{amssymb}
\usepackage{epstopdf}

\newcommand{\be}{\begin{eqnarray}}
\newcommand{\ee}{\end{eqnarray}}
\newcommand{\ba}{\begin{array}}
\newcommand{\ea}{\end{array}}

\makeatletter
\def\captionof#1#2{{\def\@captype{#1}#2}}
\makeatother
\begin{document}

\title{Multichannel matched filtering for spherical gravitational wave 
antennas}
\author{Carlos Filipe Da Silva Costa$^{(1)}$, Stefano Foffa$^{(2)}$ and 
Riccardo Sturani$^{(3,4)}$}
$(1)$ D\'epartment de Physique Nucl\'eaire et Corpusculaire, Universit\'e de
Gen\`eve\\
$(2)$ D\'epartment de Physique Th\'eorique, Universit\'e de Gen\`eve, CH-1211 Geneva, Switzerland\\
$(3)$ Istituto di Fisica, Universit\`a di Urbino, I-61029 Urbino, Italy\\
$(4)$ INFN, Sez. Firenze, Sesto Fiorentino, I-50019 Italy
\ead{Filipe.DaSilva@unige.ch,\ Stefano.Foffa@unige.ch,\ Riccardo.Sturani@uniurb.it}

\date{$30^{\rm th}$ April 2009}

\begin{abstract}
We study the performance of a multidimensional matched filter as 
a follow-up module of the coherent method recently developed
by two of us for the detection of gravitational wave bursts by
spherical resonant detectors.
We have tested this strategy on the same set of injections used
for the coherent method and found that the matched filter sensibly improves
the determination of relevant parameters as the arrival time, amplitude,
central frequency and arrival direction of the signal.
The matched filter also improves the false alarm rate, reducing it
roughly by a factor of 3.
The hierarchical structure of the whole analysis pipeline allows to obtain 
these results without a significant increase of the computation time.
\end{abstract}

\maketitle

\section{Introduction}
Whether the first direct gravitational wave (GW) detection will be made
during the forthcoming joint LIGO-Virgo \cite{intf_web} run, or rather will
be claimed by the next generation of interferometers, the intriguing era of
gravitational wave astronomy is in any case expected to begin.

In this perspective, spherical resonant detectors, having the capability of 
locating on their own the gravitational wave source in the sky, may represent 
an important resource for this arising discipline.
Contrarily to an interferometer, a sphere is a multichannel detector capable
in principle to determine all the four parameters characterizing a GW
(two angles for the arrival direction, and the amplitudes of the two
polarizations);
in order to deal with this distinguishing feature, several
data analysis issues must be however tackled (see \cite{Wagoner}-
\cite{Gottardi:2006gn} for an incomplete but elucidating historical
perspective).

In \cite{Foffa:2008pm} and \cite{Foffa:2008fd}
a coherent burst search algorithm has been proposed,
that fully exploits the multichannel
capabilities of the sphere by transforming the detector's outputs into a
five-dimensional time series corresponding to the spheroidal components of
the gravitational wave. This time series is in turn processed by a coherent,
wavelet-based module \cite{Mal} (henceforth CWS, as Coherent Waveburst for the 
Sphere) generating a list of candidate burst events endowed with all the relevant
parameters (arrival time, arrival direction, amplitude, etc...) and is also 
able, to some extent, to reject non-transverse excitations 
(which are surely not due to GW's), thus reducing the false alarm rate by a 
factor of about 10.

The reason of using wavelets rather than the traditional matched filter(MF)-ing
technique is two-fold. MF-based searches are computationally expensive, as 
they require the exploration of a parameter space which is made extremely 
large by the presence of several channels (for example, the 
multichannel MF is direction dependent, differently from the traditional
single channel one). Moreover wavelet-based searches, contrarily to a MF one,
do not require any detailed prior knowledge of a signal, a very welcome feature
for an unmodeled burst search such as the one developed in 
\cite{Foffa:2008pm,Foffa:2008fd}.\\
On the other hand, as the use of the MF is known to give the optimal 
signal-to-noise ratio (SNR) \cite{Maggiore:1900zz}, the wavelets flexibility 
has a cost in term of accuracy in the determination of the burst parameters.

To this respect, the present work can be considered as a natural evolution of 
\cite{Foffa:2008pm,Foffa:2008fd} in that it implements a multidimensional coherent MF
as a follow-up of the method cited above. This hierarchical strategy
is crucial in reducing the parameters space because for each trigger only a
little region is explored around the values indicated by the CWS
module, thus making the whole procedure reasonably fast.

The plan of the paper is the following:
in section \ref{apip} the whole pipeline is briefly described,
showing in particular how the MF module is interfaced to the previous part
of the analysis chain;
then in section \ref{resu} the improved method is tested against the
same mock data sets (software injections of signals on a background
endowed with both Gaussian and non-Gaussian noise, generated on the basis of a 
numerical simulation of the relevant components of the detector), thus showing 
the improvements obtained in the parameters estimation. 
Finally, section \ref{conc} contains our conclusions.

\section{Description of the analysis pipeline}\label{apip}
\subsection{From the transducers' outputs to the event candidates}
The first part of the pipeline, which is exhaustively
discussed in \cite{Foffa:2008pm} and \cite{Foffa:2008fd},
will be just briefly summarized here, leaving aside many technical details.
The first step consists in converting the six transducers' outputs $I_k$
into five channels corresponding to the 5 quadrupolar modes $h_N$ of the
GW tensor $h_{ij}$:
\begin{eqnarray}\label{fromItoh}
\hspace{4cm} I_k(t)\longrightarrow h_N(t)\,,
\end{eqnarray}
with
\begin{eqnarray}\label{defhY}
h_{ij}=\sum_{N=0,4}{\cal Y}^{(2)}_{m_N, ij} h_N\,,\quad\quad
Y^{(2)}_{m_N}=\sum_{i,j=x,y,z}{\cal Y}^{(2)}_{m_N, ij}n^i n^j\,,
\end{eqnarray}
where $m_N=0,1c,1s,2c,2s$ labels the components of the
real spherical harmonics $Y^{(2)}_{m_N}$ (defined and normalized as in 
\cite{Zhou:1995en}), ${\cal Y}^{(2)}_{m_N,ij}$ denotes the Cartesian component 
of the tensor spherical harmonics and $n^i$ is the versor of the arrival 
direction of the GW.

This conversion allows to extract the information about the GW from
the detector's outputs (or, equivalently, to express its noisy content
in term of its ``GW-equivalent") and is possible once a mathematical model for 
the detector is available.
In the case at hand, the model consists of a set of coupled oscillators
describing the dynamics of the relevant sphere vibrational modes, of the
transducers, and of the electrical circuit that are at the core of the 
readout devices (see \cite{Foffa:2008fd} for details).

The next step is to combine the Fourier transforms of the five $h_N$ channels
into a single channel $H$:
\begin{eqnarray}\label{fromhtoH}
\hspace{4cm} h_N(\omega)\longrightarrow H(\omega)\,.
\end{eqnarray}
This process is performed by extracting from the dataset the spectral density
matrix, defined from
\begin{eqnarray}\label{Shdef}    
S^h_{NN'}(\omega)\delta(\omega-\omega')\equiv<h_N(\omega) h^*_{N'}(\omega')>\,,
\end{eqnarray}
and then by building the following quantity:
\begin{eqnarray}\label{scacha}
H(\omega)\equiv(h\cdot h)_S\equiv
\sum_{N,N'}h^{\dag}_N(\omega)\cdot(S^h)^{-1}_{NN'}(\omega)\cdot h_{N'}(\omega)
\,.
\end{eqnarray}

This quantity is then fed into the CWS algorithm, which is a modified version 
of the Coherent Waveburst algorithm developed for LIGO 
(\cite{Klim}-\cite{Klimenko:2008fu}), that has been adapted to our
computational needs (a standard MacPro machine with two 3GHz processors)
and to our detector configuration.
This algorithm provides a list of triggers containing the various
event features (central time of events, amplitude, arrival 
direction, central spectral frequency and spectral frequency spread).

The trigger list is finally further refined through an independent
determination of the arrival direction (the so-called determinant method),
and a trigger is vetoed when the two reconstructed directions are not
compatible. To summarize:
\begin{eqnarray}\label{fromHtolist}
\hspace{1cm}H(\omega)
\longrightarrow
\{{\rm triggers}\}
\longrightarrow \{{\rm event\ candidates}\}\,.
\end{eqnarray}

\subsection{Follow-up of the events via multidimensional
matched filtering}\label{mfilpipe}
If the time dependence, arrival direction and polarization
of the GW signal were known, that is if the expression for the five functions 
$h^{sig}_N(\omega)$ were available, the optimal detection strategy would be to 
perform a multidimensional matched filter.
As explained in \cite{Stevenson:1996rw}, the multichannel MF consists in
analyzing the following scalar channel:
\begin{eqnarray}\label{multiMF}
H_{MF}(\omega)\equiv(h^{sig}\cdot h)_S\, ,
\end{eqnarray}
and is the strategy which gives the optimal SNR
\begin{eqnarray}\label{SNRMF}
SNR=\frac{\int (h^{sig}\cdot h)_S {\rm d}\omega}{\sqrt{\int (h^{sig}\cdot h^{sig})_S{\rm d}\omega}} 
\end{eqnarray}
for the given signal $h^{sig}$. The integrals are done on both positive and 
negative frequencies.
However, in a real experimental situation $h^{sig}$ is not known a priori
and to employ efficiently a MF one should explore a huge number of points in a 
parameter space describing the temporal profile of the signal, as well as the 
arrival direction and polarization. Thus the use of a MF in such a context is 
computationally very demanding.

The CWS method exposed in the previous paragraph
has been shown to find candidate events with a good efficiency/reliability
ratio, without need to know the form of the signal.
This allows to keep under control the computation time but of course it
comes at a price for what concerns parameter estimation.

The improvement we are introducing in this work consists in performing a
multidimensional MF only in the time segments where candidates events have
been selected by the CWS, and to restrict the analysis to a small portion
of parameter space centered around the parameters provided by the CWS triggers.

Specifically, we consider a two parameter family of sine-Gaussian shapes
to describe the time profile of the signal
\begin{eqnarray}\label{singau}
h^{sig}_N(t|t_0,\Delta,f_0,\theta,\phi)=
h_0{\rm e}^{-(t-t_0)^2/(2\Delta^2)}\sin\left[2 \pi f_0 (t-t_0)\right] 
Y_{m_N}^{(2)}(\theta,\phi)\,,
\end{eqnarray}
impinging from the generic direction characterized by polar angles 
$\theta,\phi$.
Since we have verified that the unmodeled search alone already gives a good
estimation of the frequency spread $1/(2 \pi\Delta)$
(an estimation which is not improved by the MF procedure),
we kept it as fixed, in such a way that,
for each candidate event we are basically exploring a small portion of the
4-dimensional parameter space $\{t_0,f_0,\theta,\phi\}$.

Such a space is explored through a maximum-seeking algorithm, which searches
for the point characterized by the maximum SNR. 

We maximized first on the arrival time, as the peak of the SNR is very sharp
against this parameter, then on the central frequency and finally over the two 
angles of the arrival direction together, keeping at each stage the
other parameters fixed at the value indicated by the CWS triggers. This is a 
simplified choice of course, with respect to building a full parameter 
template bank, for instance, but it allows to obtain good results while 
keeping a low computational burden.

Once found, the maximum-SNR point in the parameter space defines the improved
parameters of the candidate event, and if they happen to be too much different 
from the ones given by the unmodeled search, the event is discarded as 
spurious.

This means that the MF follow-up procedure, as well as improving the events' 
parameters, acts as a further veto against non-GW triggers, i.e. against
signals which are not compatible with the transverse-traceless nature of GW's. 
Such false triggers can come from Gaussian noise or from non-Gaussian disturbancies.
The CWS method alone is already optimized to reduce the latter's, and we will show
that the application of the MF follow-up improves the performance with respect to the
rejection of false triggers due to Gaussian noise.

\section{Results}\label{resu}

We have tested the MF method on the triggers obtained in \cite{Foffa:2008fd}, 
i.e. 50 GW-injections, (25 for each polarization) for each of the four $h_{rss}$
amplitudes: $3,5,7,10\times 10^{-21}{\rm Hz^{-1/2}}$ and also on an equal 
number of non-GW injections (signals whose power has been
distributed randomly among the five sphere channels)
with the same amplitude.
The injected signals are sine-Gaussian of the same type of the matching function given in 
(\ref{singau}), this allows us to assess the precision in parameter
estimation irrespectively of a mismatch between the filter and the signal in 
the data. An eventual mismatch could of course worsen the estimations,
but we do not investigate here on that.

For the sake of clarity, we have reported in table \ref{GWnonGW} the detection efficiency for
both GW and non GW-injections obtained in \cite{Foffa:2008fd}; these numbers, which show that the CWS
alone already reduces by a factor of 5-10 the number of non-Gaussian false triggers,
have not changed in the present analysis.
 
\begin{table}
  \begin{center}
    \begin{tabular}{|c|c|c|}
      \hline
      $h_{rss} ({\rm Hz^{-1/2}})$ & \#GW & \#non-GW\\
      \hline
      $10^{-20}$ & 48 & 8\\ 
      \hline
      $7\cdot10^{-21}$ & 33 & 5\\ 
      \hline
      $5\cdot10^{-21}$ & 30 & 7\\ 
      \hline
      $3\cdot10^{-21}$ & 24 & 9\\ 
      \hline
    \end{tabular}
    \caption{Detection efficiency for injected GW signals and non-GW
      disturbances. Adapted from \cite{Foffa:2008fd}.}
    \label{GWnonGW}
  \end{center}
\end{table}

\subsection{Efficiency vs. false alarm rate}

The GW detection efficiency alone is not a satisfactory measure of the
goodness of the analysis method, as it must be completed by informations about
the false alarm rate.
A Receiver Operation Characteristic (ROC) curve has been constructed in order to find
the set of analysis parameters (the most important being in our case the maximal
allowed distance between the two direction reconstruction methods, $\delta_{thr}$)
that provides the optimal efficiency vs. false alarm performance.

\begin{center}
  \begin{figure}[htbp]
    \includegraphics[width=.9\linewidth]{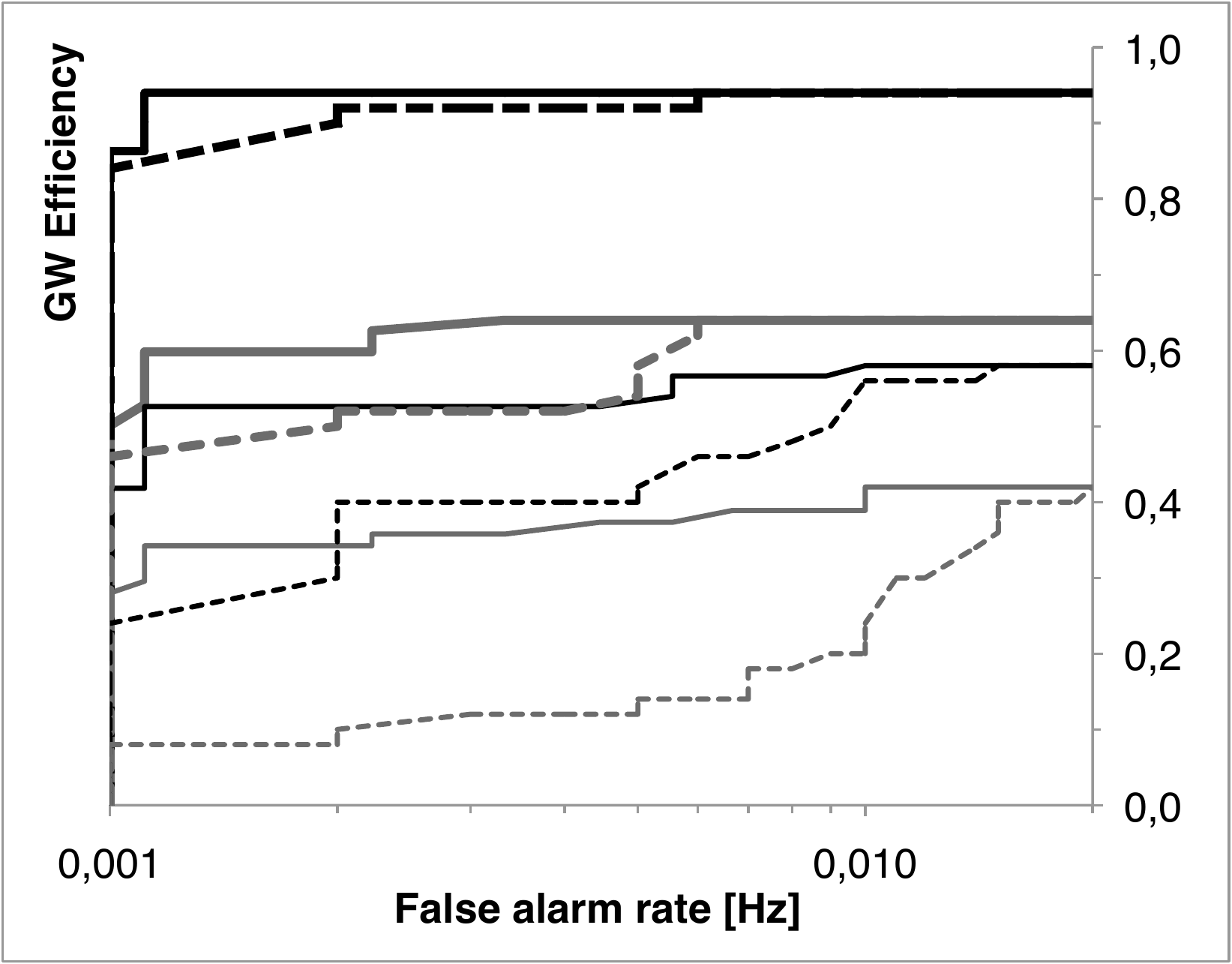}
    \caption{ROC curves (detection efficiency vs. false alarm rate) for the 
      GW-injections, obtained by varying the maximum allowed distance between
      the direction reconstruction methods. 
      Broken lines show the ROC obtained in \cite{Foffa:2008fd} 
      with the unmatched method, varying the distance between the CWS and
      the determinant method; solid lines show the results after the 
      application of the matched filter and are obtained varying the distance 
      between CWS and MF methods.
      Lines of increasing
      thickness correspond to injections sets of increasing amplitude, 
      $h_{rss}=\{3,5,7,10\}\times 10^{-21}{\rm Hz}^{-1/2}$.
}
    \label{roc}
  \end{figure}
\end{center}

In figure \ref{roc} we report the ROC curves both for the 
CWS unmodeled search and after the application of the MF, for different injection strengths.
The false alarm rate is obtained from the analysis of simulated data containing Gaussian noise only.
The figure shows that, at a given level of the efficiency, the MF method allows to
reduce considerably the false alarm rate (by roughly a factor of 3 for $\delta_{thr}=0.2$ radians,
which the same working point chosen in \cite{Foffa:2008fd}).

\newpage
\subsection{Parameters estimation}
\subsubsection{Arrival time}
The MF improves by an order of magnitude the arrival time
estimation, as can be seen in figure \ref{time_rec}.
The results are summarized in table \ref{sum_time},
where we plotted the average and the standard deviation
of the quantities $t_{CWS}-t_{inj}$ and $t_{MF}-t_{inj}$.
\begin{minipage}[t]{1\linewidth}
  \begin{center}
    \includegraphics[width=.49\linewidth,angle=0]{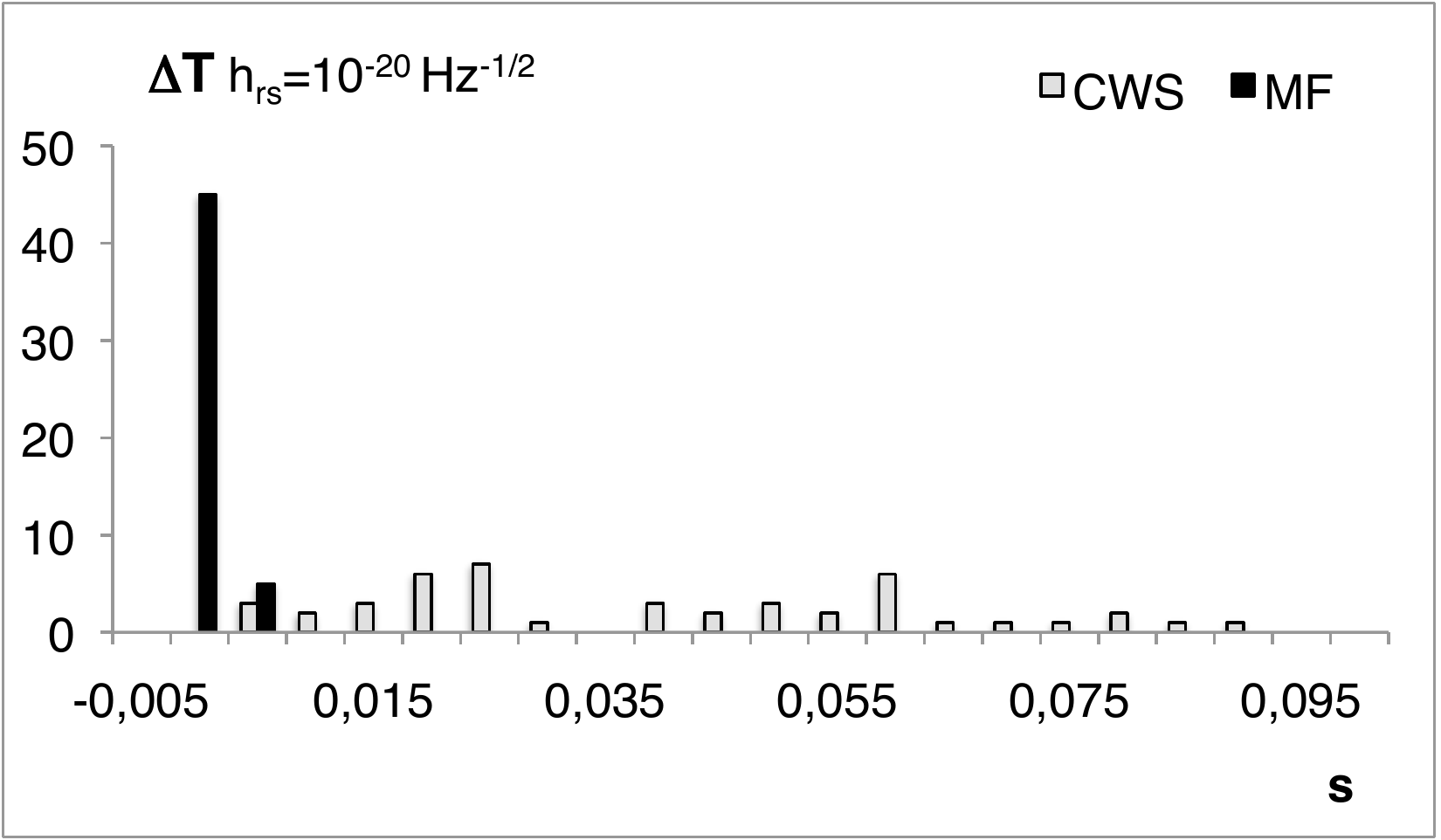}
    \includegraphics[width=.49\linewidth,angle=0]{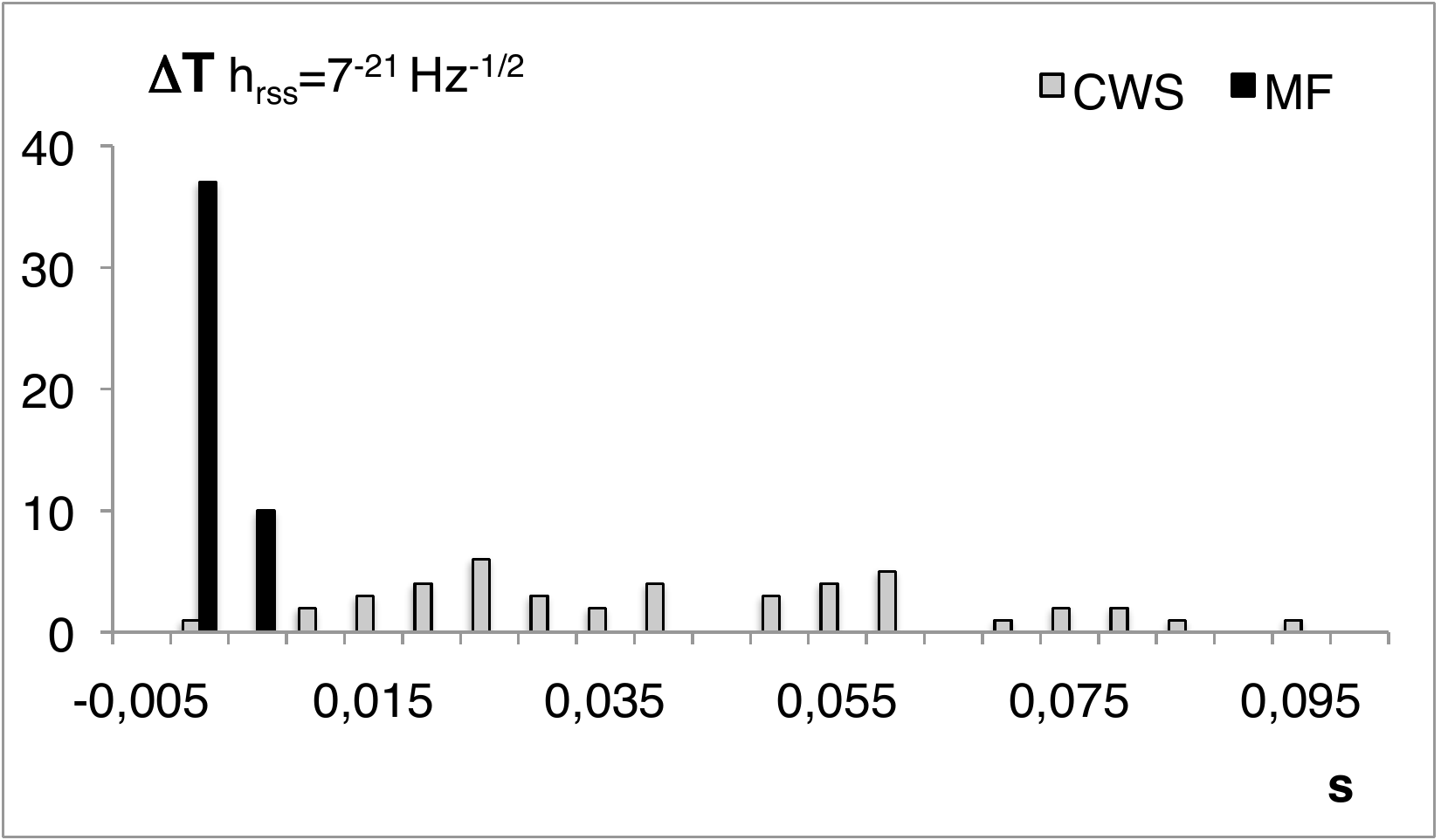}
    \includegraphics[width=.49\linewidth,angle=0]{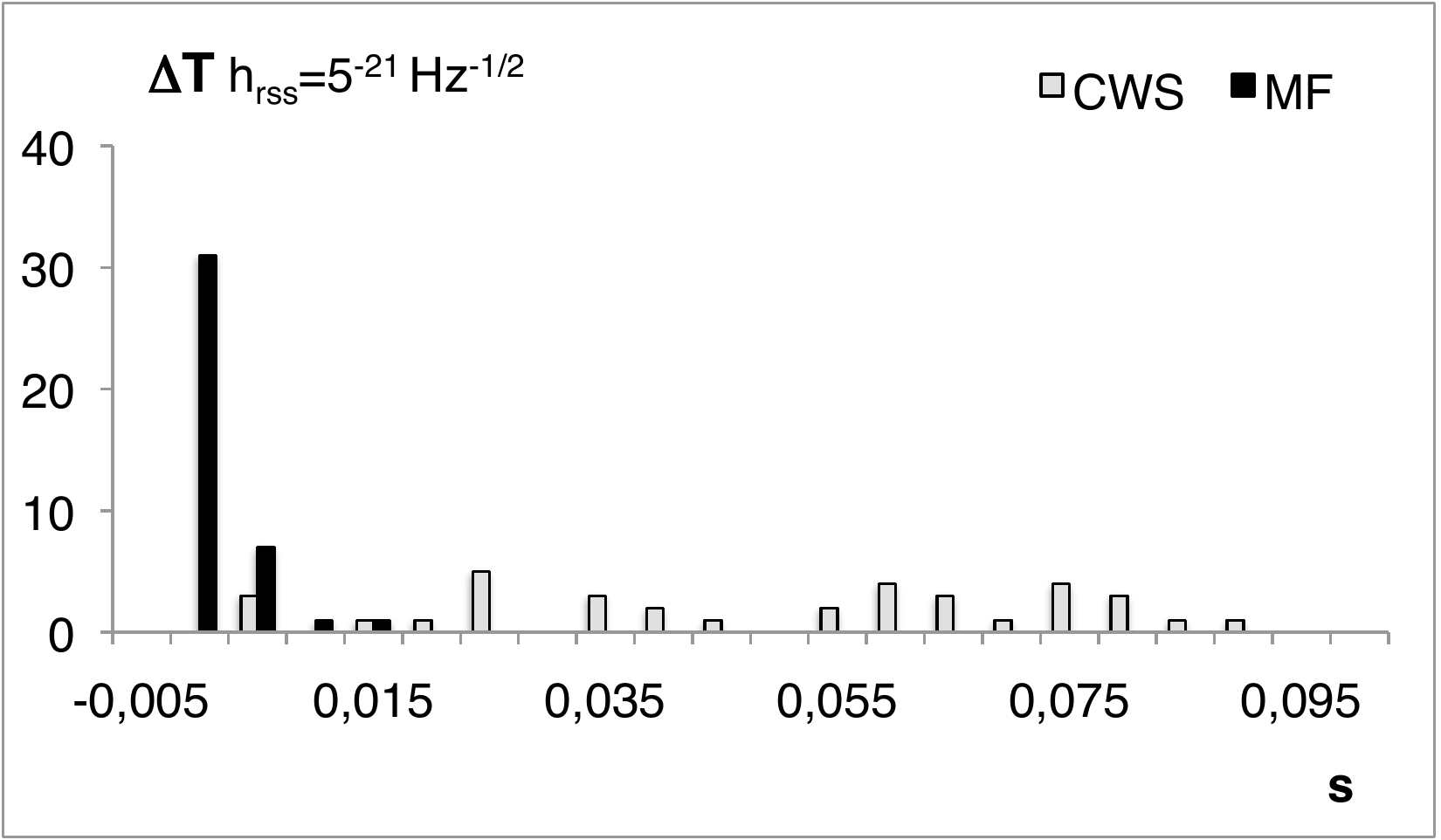}
    \includegraphics[width=.49\linewidth,angle=0]{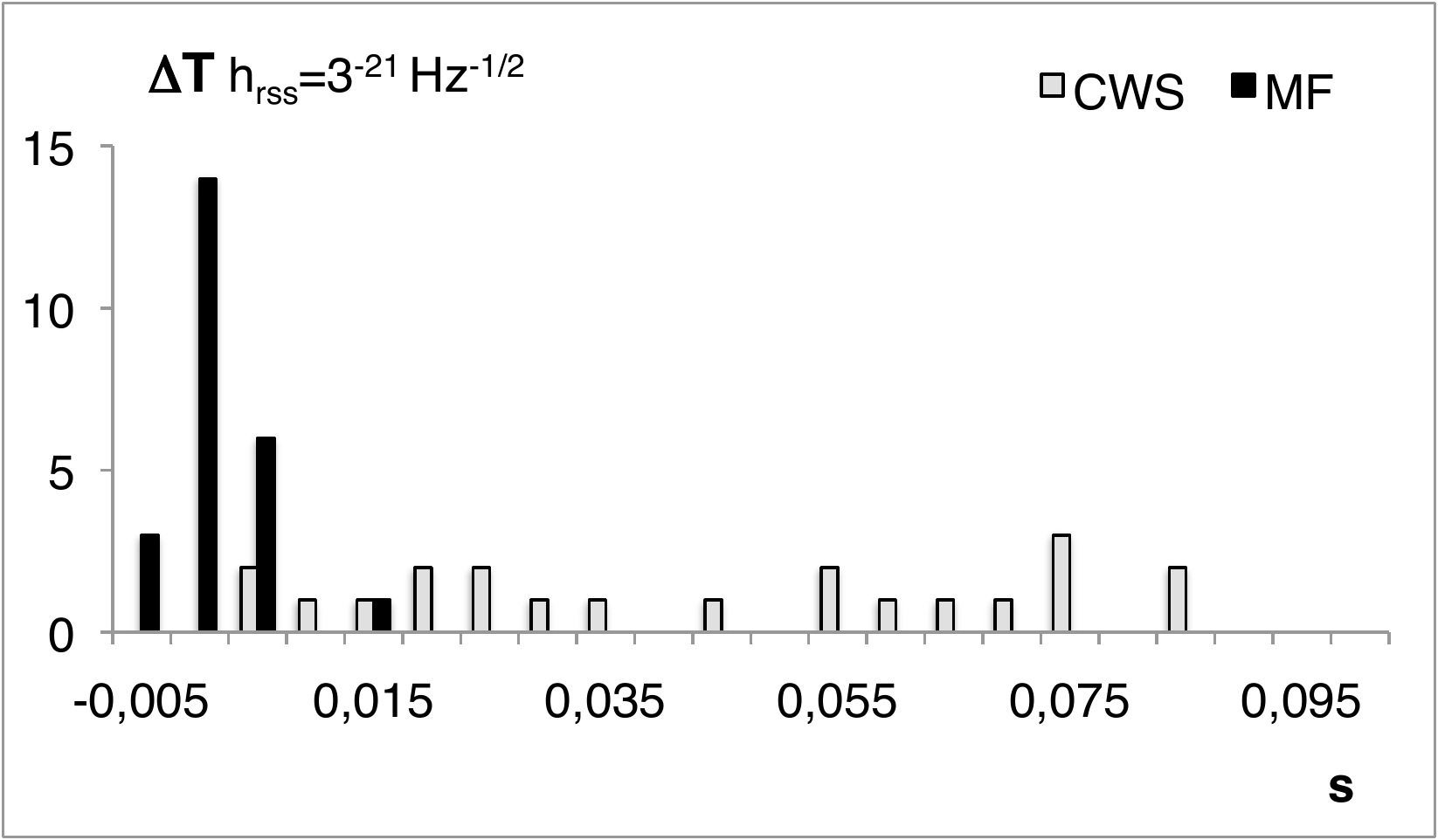}
    \captionof{figure}{
      \caption{Histograms showing the arrival time reconstruction precision
        for the unmodeled (light) and matched-filtered (black) search, at different
        amplitudes.}
      \label{time_rec}
    }
    \vspace{.5cm}
    \begin{tabular}{|c|c|c|c|c|}
      \hline
      $h_{rss} ({\rm Hz^{-1/2}})$ &\multicolumn{2}{|c|}{$\Delta t_U$ (msec)} & 
      \multicolumn{2}{|c|}{$\Delta t_{MF}$ (msec)} \\
      \cline{2-5}
      & $\mu$ & $\sigma$ & $\mu$ & $\sigma$\\
      \hline
      $10^{-20}$ & 43 & 28 & -1.2 & 2.1\\ 
      \hline
      $7\cdot10^{-21}$& 43 & 25 & -0.48 & 2.9\\ 
      \hline
      $5\cdot10^{-21}$ & 52 & 28 & -0.24 & 3.4\\ 
      \hline
      $3\cdot10^{-21}$& 48 & 29 & -0.40 & 4.7\\ 
      \hline
    \end{tabular}
    \captionof{table}{
      \caption{Average $\mu$ and standard deviation $\sigma$
        of the difference
        (in milliseconds) between the arrival time determination and the actual
        injected one, for the unmodeled search (U) and for the matched filter.
        The fact that $|\mu|<\sigma$ in the MF case means that our estimator
        is unbiased, and in such case the precision of the method is given
        by $\sigma$.}
      \label{sum_time}
    }
  \end{center}
\end{minipage}

\newpage
\subsubsection{Central frequency}
Also the estimation of $f_0$ benefits of an one order of magnitude improvement
with the MF method, see table \ref{sum_freq}.
More specifically, most of the improvement comes from the correction of a
few very imprecise estimations given by the CWS, as can be seen in the
histograms of figure \ref{freq_rec}.
\begin{minipage}[b]{1\linewidth}
  \begin{center}
    \includegraphics[width=.49\linewidth,angle=0]{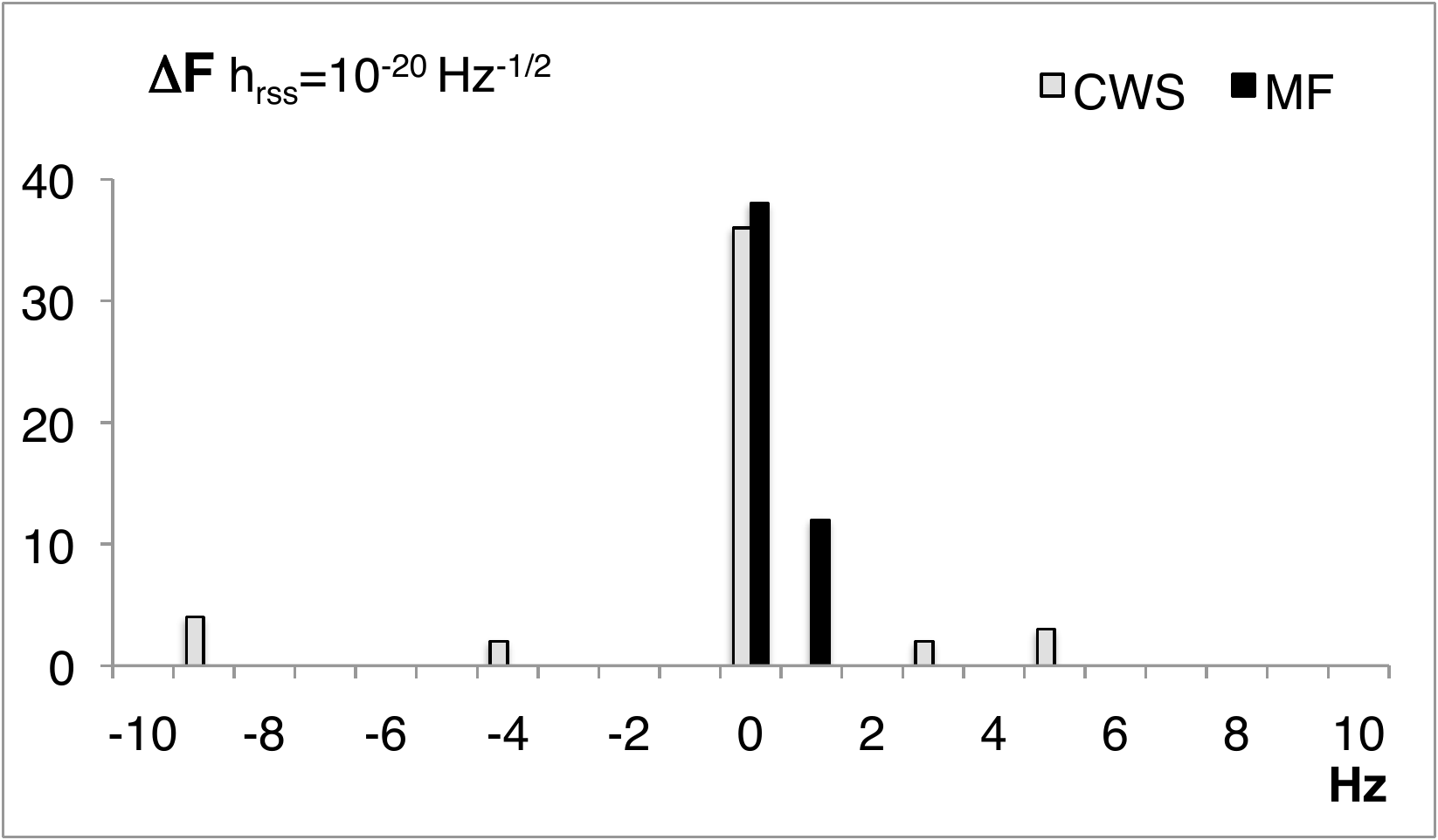}
    \includegraphics[width=.49\linewidth,angle=0]{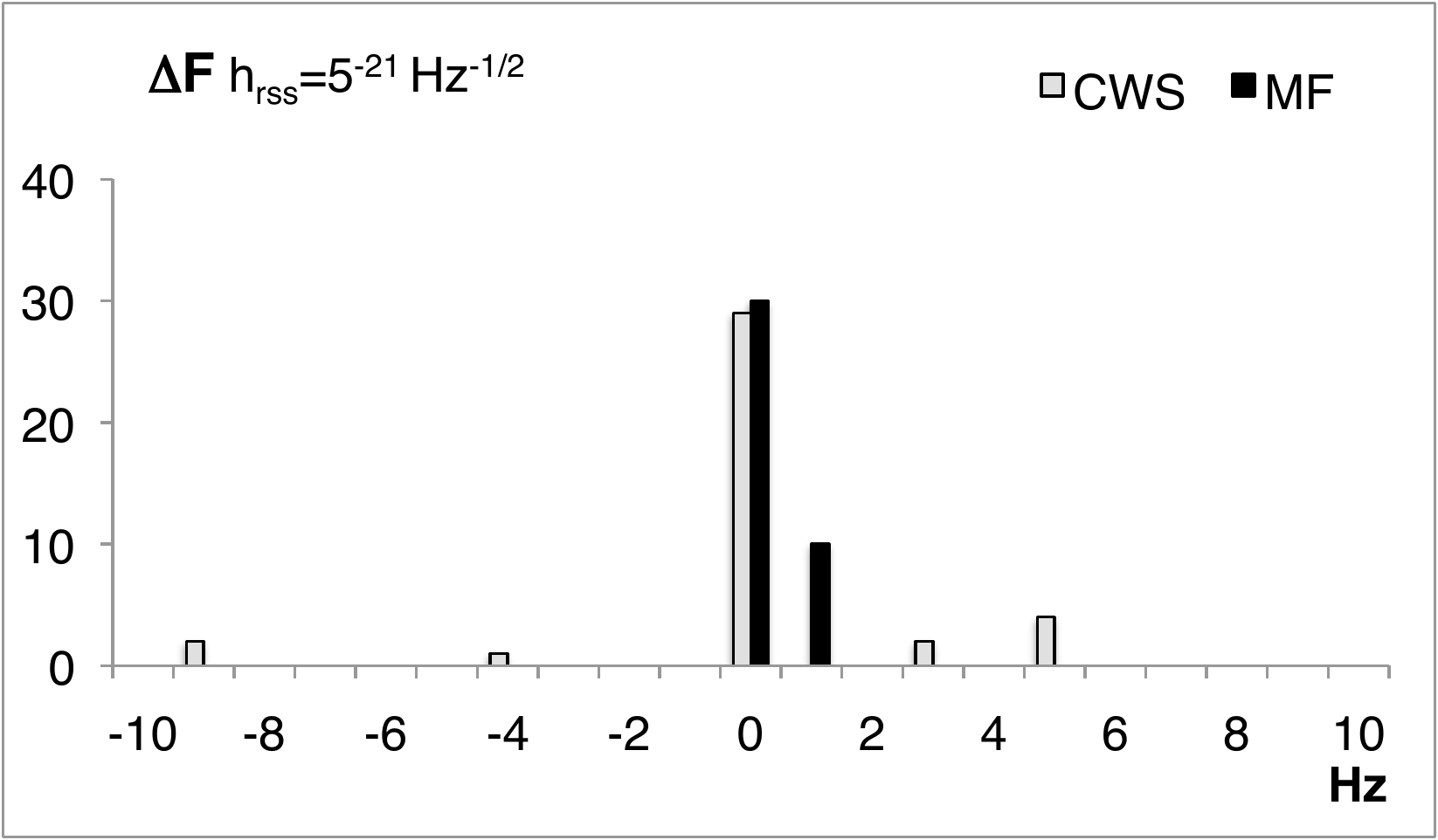}
    \includegraphics[width=.49\linewidth,angle=0]{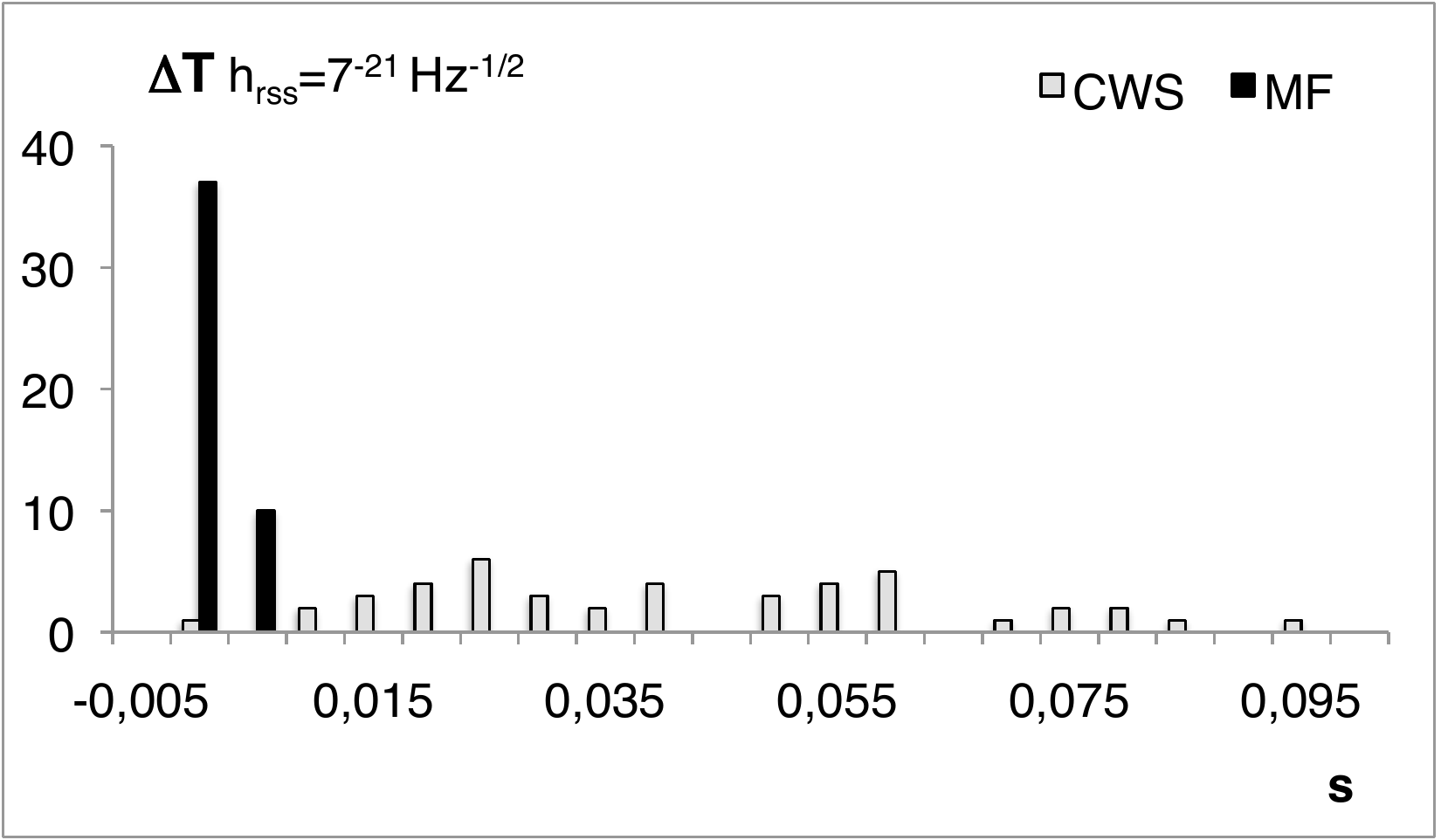}
    \includegraphics[width=.49\linewidth,angle=0]{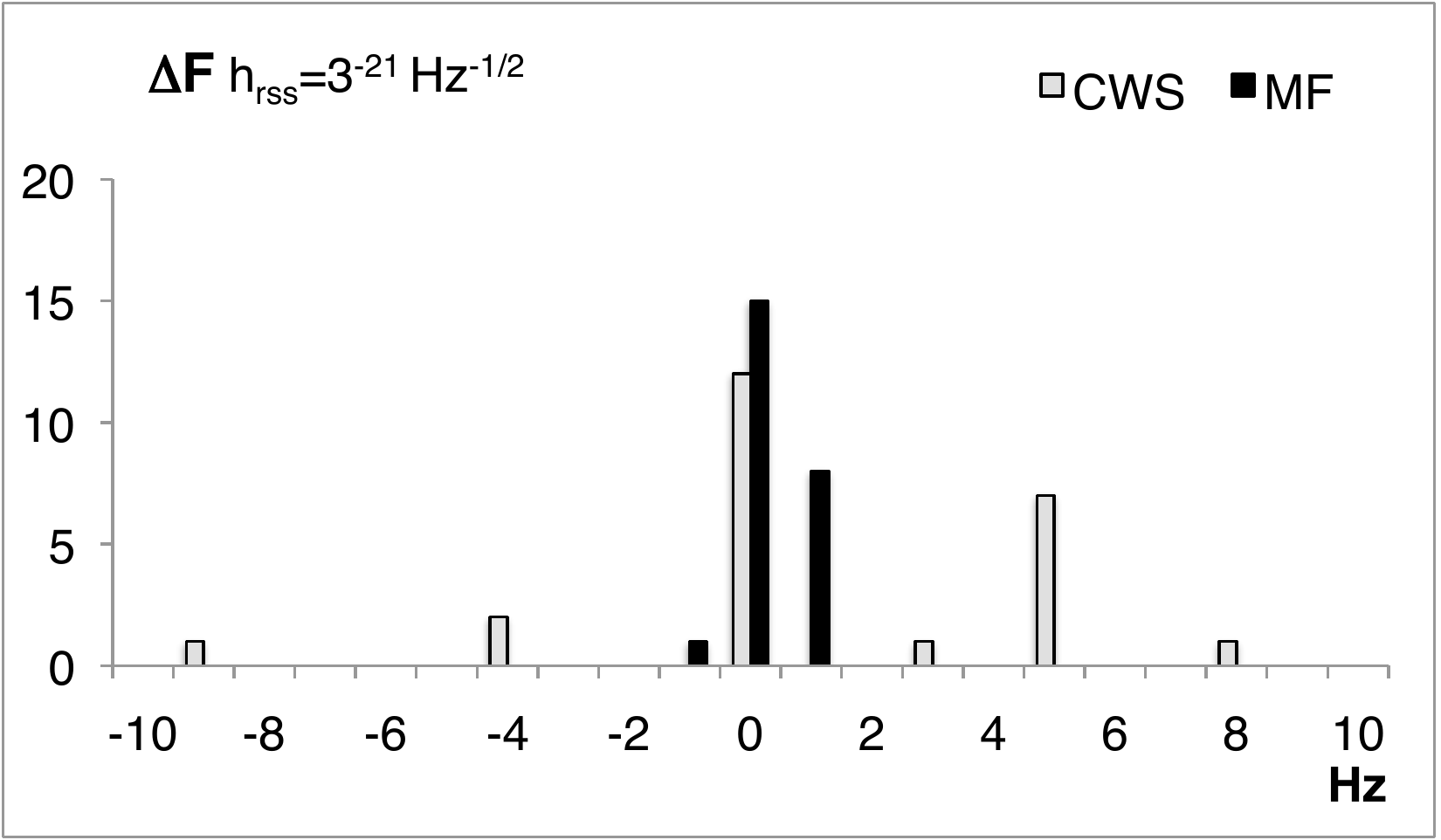}
    \captionof{figure}{
      \caption{The four histograms show the distribution of the quantities
        $f_0^{CWS}-f_0^{inj}$ (light) and $f_0^{MF}-f_0^{inj}$ at different
        injected amplitudes.}
      \label{freq_rec}
    }
    \vspace{.5cm}
    \begin{tabular}{|c|c|c|c|c|}
      \hline
      $h_{rss} ({\rm Hz^{-1/2}})$ &\multicolumn{2}{|c|}{$\Delta f_0^U$ (Hz)} & 
      \multicolumn{2}{|c|}{$\Delta f_0^{MF}$ (Hz)} \\
      \cline{2-5}
      & $\mu$ & $\sigma$ & $\mu$ & $\sigma$\\
      \hline
      $10^{-20}$ & 0.28 & 4.6 & 0.08 & 0.28\\ 
      \hline
      $7\cdot10^{-21}$ & 0.60 & 4.7 & 0.08 & 0.32\\ 
      \hline
      $5\cdot10^{-21}$ & 0.70 & 4.1 & 0.10 & 0.27\\ 
      \hline
      $3\cdot10^{-21}$& 0.97 & 3.6 & 0.06 & 0.54\\ 
      \hline
    \end{tabular}
    \captionof{table}{
      \caption{Average $\mu$ and standard deviation $\sigma$ of $\Delta f_0$
        (in Hz) for the unmodeled search (U) and for the matched filter.
        The fact that $|\mu|<\sigma$ means that our estimator is unbiased,
        and in such case the precision of the method is given by $\sigma$.} 
      \label{sum_freq}
    }
  \end{center}
\end{minipage}

\newpage
\subsubsection{Arrival direction}
In fig.~\ref{dir_rec} the direction reconstruction of the unmodeled and of
the MF search are plotted.
In this case the improvement, although clearly visible,
is not as spectacular as in the previous cases,
and appears to be more marked for $\phi$ than for $\theta$; this last feature
might depend on the chosen injection direction, and further study is
necessary to understand the behavior for generic arrival directions.
Quantitative results for the angular distance $\delta s$
are summarized in tab.~\ref{sum_dir}.
\begin{minipage}[b]{1\linewidth}
  \begin{center}
    \includegraphics[width=.45\linewidth,angle=0]{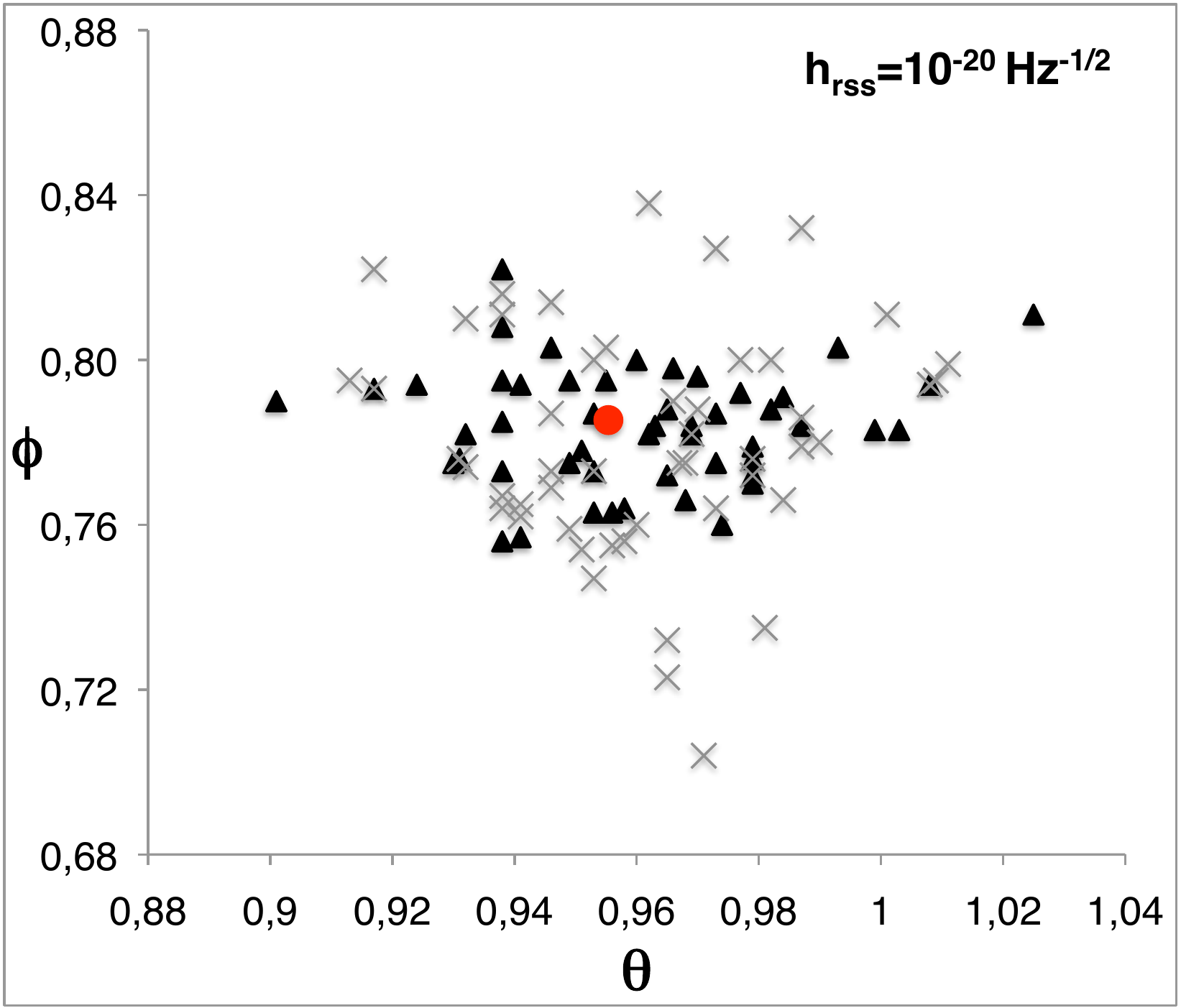}
    \includegraphics[width=.45\linewidth,angle=0]{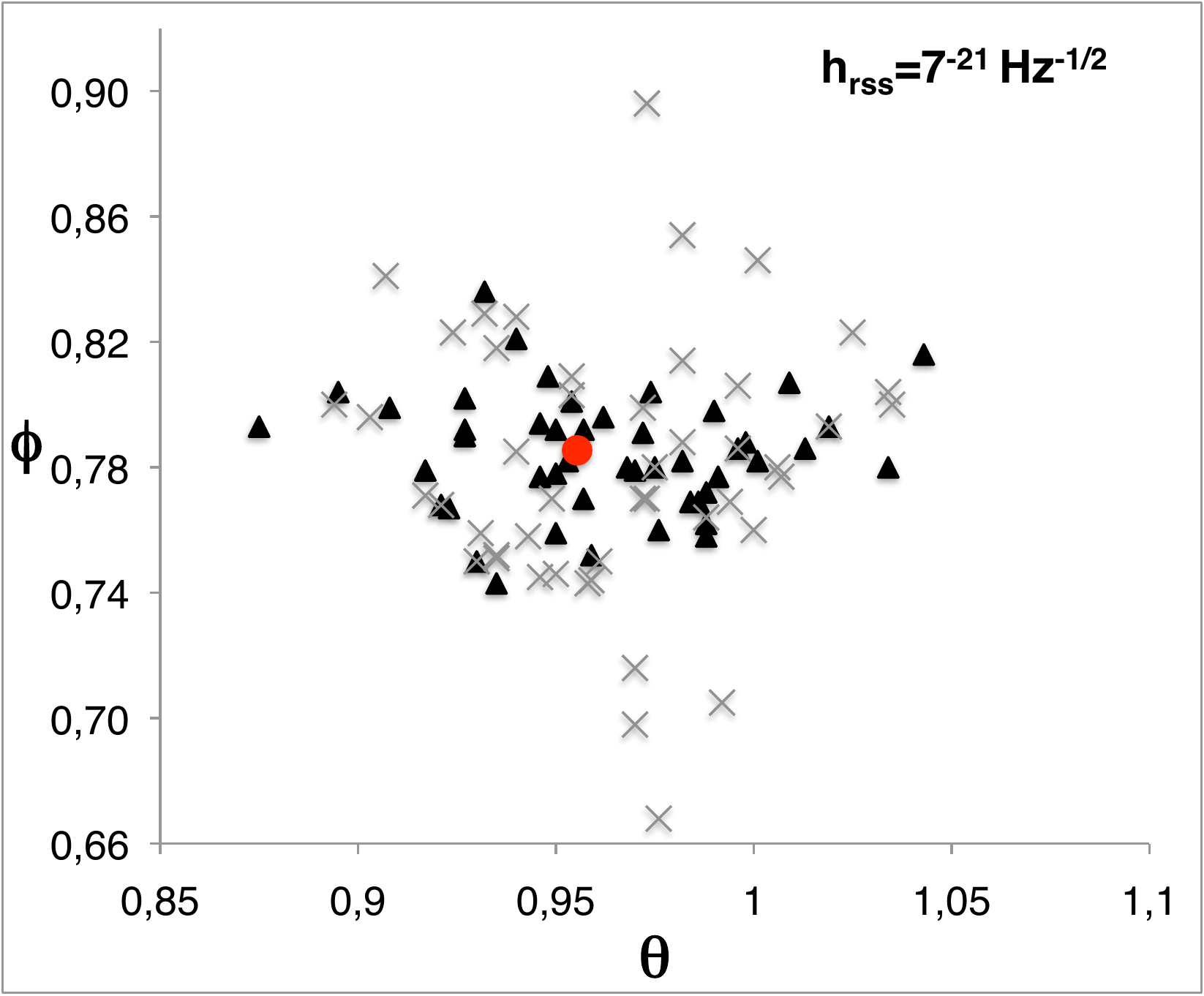}
    \includegraphics[width=.45\linewidth,angle=0]{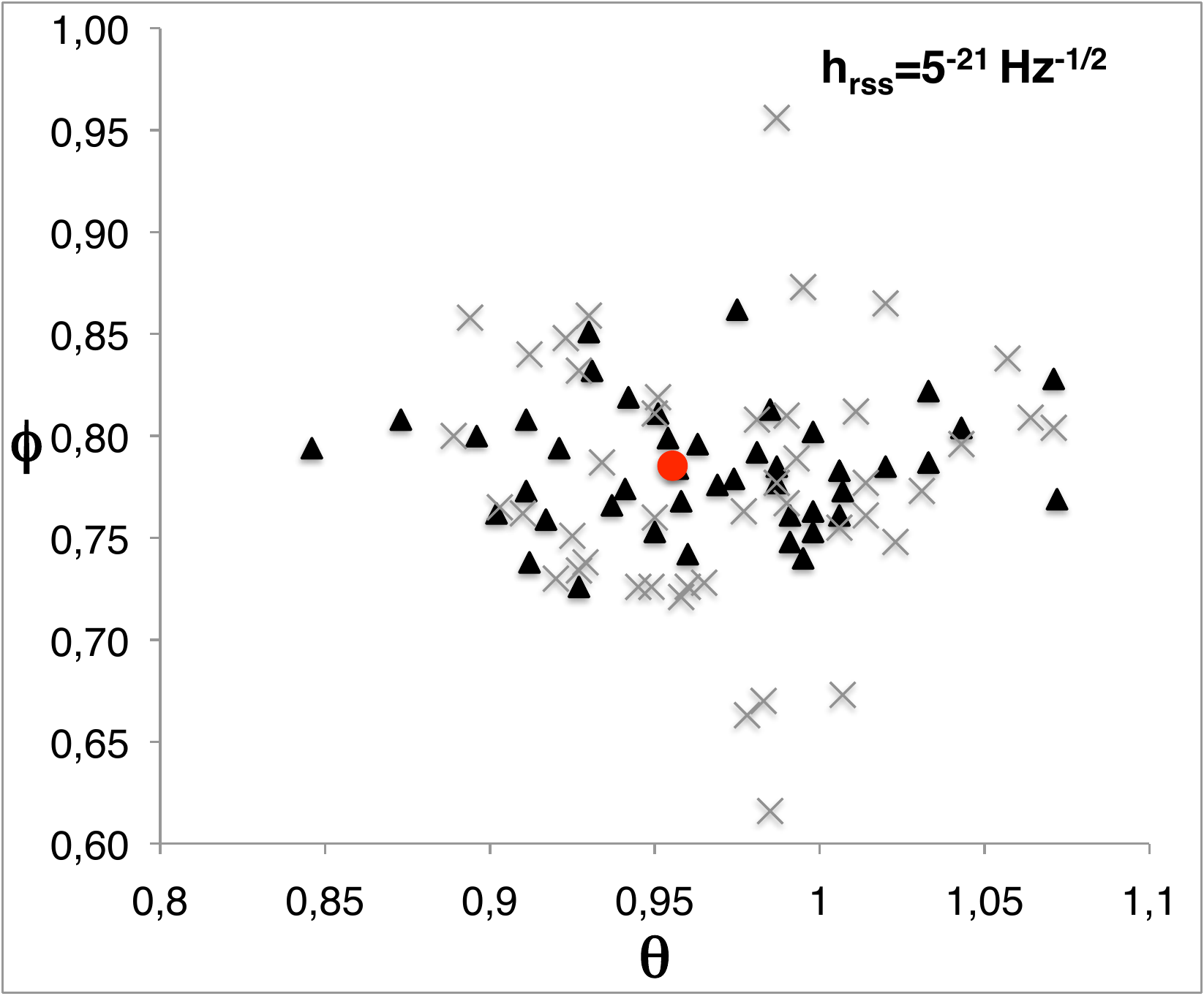}
    \includegraphics[width=.45\linewidth,angle=0]{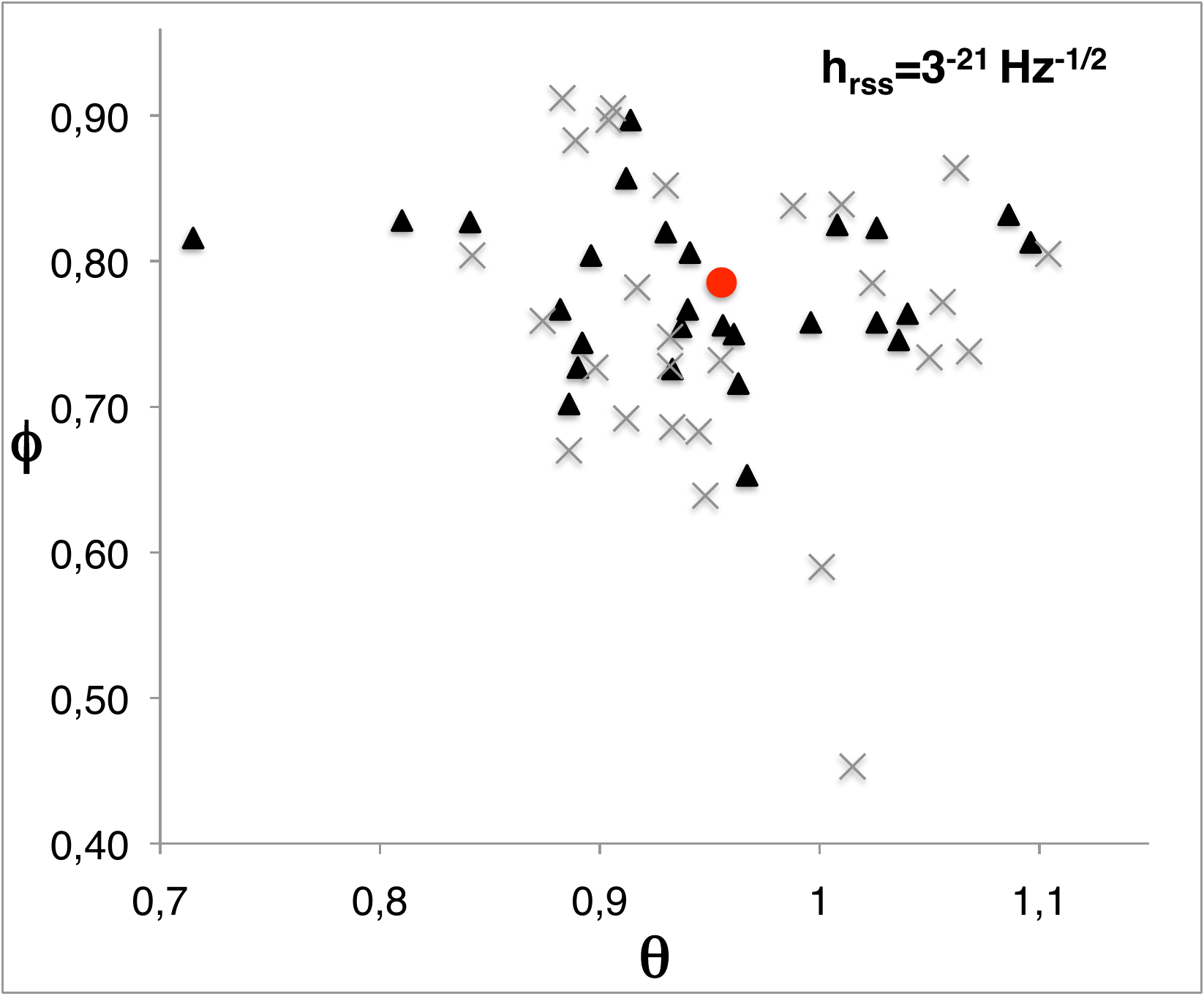}
    \captionof{figure}{
      \caption{Arrival direction reconstruction by the unmodeled search
        (crosses) and by the MF (triangles) for the four given signal
        amplitudes. The circle represents the injection direction.}
      \label{dir_rec}
    }
    \vspace{.5cm}
    \begin{tabular}{|c|c|c|}
      \hline
      $h_{rss} ({\rm Hz^{-1/2}})$ & $\delta s_U$ & $\delta s_{MF}$ \\
      \hline
      $10^{-20}$ & 0.030 & 0.024 \\ 
      \hline
      $7\cdot10^{-21}$ & 0.044 & 0.034 \\ 
      \hline
      $5\cdot10^{-21}$ & 0.58 & 0.050 \\ 
      \hline
      $3\cdot10^{-21}$ & 0.092 & 0.083 \\ 
      \hline
    \end{tabular}
    \captionof{table}{
      \caption{Average angular distance (in radians)
        between the direction determination and the actual injection direction,
        for the unmodeled search (U) and for the matched filter.
        Differently from the previous cases, the angular distance is clearly a
        biased parameter (because it is not distributed around zero,
        by definition), so the accuracy of the estimation is well described by
        the average.} 
      \label{sum_dir}
    }
  \end{center}
\end{minipage}

\subsubsection{Amplitude}
The CWS method have not been explicitly calibrated to measure
the amplitude of the signal.
However, it is still interesting to see that with the use
of the Matched Filtering the relative precision in the determination is a few 
percents, as reported in in table \ref{sum_amp}.

\begin{table}
  \begin{center}
    \begin{tabular}{|c|c|c|}
      \hline
      $h_{rss} ({\rm Hz^{-1/2}})$ & \multicolumn{2}{|c|}{$\Delta h_0^{MF}/h_0$} \\
      \cline{2-3}
      & $\mu$ & $\sigma$\\
      \hline
      $10^{-20}$ & $2.9\cdot 10^{-3}$ & 0.057\\ 
      \hline
      $7\cdot10^{-21}$ & $3.5\cdot 10^{-3}$ & 0.066\\ 
      \hline
      $5\cdot10^{-21}$ & $2.0\cdot 10^{-3}$ & 0.061\\ 
      \hline
      $3\cdot10^{-21}$ & $2.5\cdot 10^{-3}$ & 0.086\\ 
      \hline
    \end{tabular}
    \caption{Average $\mu$ and standard deviation $\sigma$ of
      $(h_0^{MF}-h_0^{inj})/h_0^{inj}$.
      The fact that $|\mu|<\sigma$ means that our estimator is unbiased,
      and in such case the precision of the method is given by $\sigma$.} 
    \label{sum_amp}
  \end{center}
\end{table}

\section{Conclusions}\label{conc}
Matched filtering techniques are widely used in gravitational wave analysis. 
However they can have a high computational cost, because of the possibly large 
volume of parameter space required by a comprehensive search.
The case of spherical detector is even worse as its multimode 
feature makes it highly sensitive to the arrival direction, adding the 
source position in the sky as further parameter.\\
Here we have shown how, in the case of burst searches,
combining an unmodeled search with a matched one 
can exploit the goodies of both while taming the baddies of both.
The unmodeled makes no assumption on the type of signal, but gives a 
non-optimal estimation of the parameters of the burst events triggered.
On the other hand the matched filtered search is optimal once the shape of the 
signal is known. Here we use the output triggers of the unmodeled search to 
massively restrict the search in the parameter of the best matching template;
more specifically we used a 4-dimensional parameter space (arrival time, 
central frequency and source angles in the sky) to present an example of 
matched filtering search. Our search is very much simplified with respect to a 
general one as the parameter space is not fully  explored.\\
To make a comprehensive analysis one would need template banks and systematic 
scanning of the parameter space, but even with this primitive attempt we have 
shown how the first estimate made by the unmodeled search can be improved
by a very light matched filter search.

\section*{Acknowledgments}
It is a pleasure to thank M. Maggiore and M. Pohl for useful discussions and 
encouragement. The work of FS is supported by the Fond National Suisse.

\end{document}